\documentclass[preprint,aps,amsmath,byrevtex,prd,titlepage,superscriptaddress,
nofootinbib
]{revtex4-2}

\usepackage{dcolumn}
\usepackage{amsmath}
\usepackage{amssymb}
\usepackage{bm}
\usepackage{hyperref}
\usepackage{graphicx}
\usepackage{slashed}
\usepackage{xcolor}

\newcommand{\beq}{\begin{equation}}
\newcommand{\eeq}{\end{equation}}
\newcommand{\eq}[1]{Eq.~(\ref{#1})}

\begin{document}

\title{
Energy levels of multiscale bound states from QED energy-momentum trace }

\author {Michael I. Eides}
\email[Email address: ]{meides@g.uky.edu}
\affiliation{Department of Physics and Astronomy,\\
University of Kentucky, Lexington, KY 40506, USA}
\author{Vladimir A. Yerokhin}
\email[Email address: ]{vladimir.yerokhin@mpi-hd.mpg.de}
\affiliation{Max Planck Institute for Nuclear Physics,\\
Saupfercheckweg 1, D 69117 Heidelberg, Germany}

\begin{abstract}
Energy levels of QED bound states, which depend on a number of independent mass parameters, can be calculated as matrix elements of the QED energy-momentum tensor trace. As an example of such system we consider muonic hydrogen. The leading one-loop corrections to its energy levels depend on the electron and muon masses. These  corrections are calculated as matrix elements of the energy-momentum tensor trace. Respective one-loop trace diagrams are different from the standard Lamb shift diagrams. We explain analytically and diagrammatically why  two different sets of diagrams lead to the same results. Similar relationships  should also hold beyond the one-loop approximation.

\end{abstract}

\maketitle


\section{Introduction}

Energy-momentum tensor (EMT) $T^{\mu\nu}$, its form factors and anomalous trace became an active field of experimental and theoretical research in the mid-1990s \cite{Ji:1996ek,Ji:1996nm,Radyushkin:1996ru,Collins:1996fb,Kharzeev:1998bz,Berger:2001xd}. EMT form factors describe mechanical properties of hadrons \cite{Polyakov:2002yz}  and their gravitational interactions \cite{Kobzarev:1962wt,Pagels:1966zza}. Both experimental and theoretical research is concentrated on the properties of hadron EMT, described by the low-energy QCD. Low-energy effective theories,  nonperturbative methods and models are used in theoretical research on hadron EMT, see, e.g.,  \cite{Ji:1994av,Ji:1995sv,Hudson:2017xug,Polyakov:2018zvc,Kharzeev:2021qkd,Liu:2021gco,
Lorce:2021xku,Ji:2020bby,Ji:2021pys,Ji:2021mtz,Metz:2020vxd} and references therein.

Quantum electrodynamics (QED) EMT is similar to the QCD EMT, but unlike QCD, QED admits low-energy perturbative calculations. Due to availability of perturbative calculations research on QED EMT  can provide a new perspective on the properties of  EMT in gauge theories. Perturbative QED calculations of free particle EMT were initiated long time ago in \cite{Milton:1971xnd,Milton:1973zz,Berends:1975ah,Milton:1976jr} and were further developed in recent papers \cite{Ji:1998bf,Kubis:1999db,Donoghue:2001qc, Rodini:2020pis,Sun:2020ksc,Ji:2021qgo,Metz:2021lqv,Ji:2021mfb,Ji:2022exr,Freese:2022jlu,
Eides:2023uox,Czarnecki:2023yqd,Czarnecki:2023dcv,Eides:2024tzh,Eides:2025,Chen:2025iul}, where a number of one-loop corrections to form factors, matrix elements and EMT trace for a free and bound electron were calculated.

Hadrons are QCD bound states, so from the perspective of  hadron physics, perturbative QED calculations of bound electron EMT are of special interest \cite{Ji:2021pys,Sun:2020ksc,Ji:2021qgo,Ji:2021mfb,Ji:2022exr,
Czarnecki:2023yqd,Czarnecki:2023dcv,Eides:2024tzh,Eides:2025}. The one-loop Lamb shift in hydrogen was calculated in \cite{Eides:2024tzh,Eides:2025}  using the well-known (see, e.g., \cite{Ji:2021qgo,Dudas:1991kj} and references therein) relationship between the EMT trace and mass (energy levels) of a particle (fundamental or a bound state) with zero momentum ($\bm p=0$) 

\beq \label{traceqmass}
E =\frac{\int d^3x\langle\bm 0|{T^\mu}_\mu(x)|\bm 0\rangle}{\langle\bm 0|\bm 0\rangle}.
\eeq

\noindent
This universal formula is valid in any quantum field theory both perturbatively and nonperturbatively.  We will use it in perturbation theory with nonrelativistic normalization.

The set of Feynman diagrams which contribute to the matrix element in \eq{traceqmass} in ordinary hydrogen does not coincide with set of the textbook diagrams  for the Lamb shift.  It was explained analytically and diagrammatically in  \cite{Eides:2024tzh,Eides:2025} why two different sets of Feynman diagrams lead to one and the same result. The calculations and interpretation of the results in \cite{Eides:2024tzh,Eides:2025}  were significantly simplified by the presence of only one mass scale, namely, the electron mass (only nonrecoil corrections were considered in \cite{Eides:2024tzh,Eides:2025}). 

In the case of bound state energy levels which depend not on one, but on a number of independent mass parameters, the argumentation presented in \cite{Eides:2024tzh} is insufficient. Our goal below is to  generalize considerations in \cite{Eides:2024tzh} to the case of a few independent mass parameters. We show that the diagrams for the EMT trace arise as logarithmic mass derivatives with respect to all independent mass parameters  of the usual  diagrams for the energy levels. It is explained why this means that the EMT trace matrix element can be used for calculations of energy levels.  We illustrate these conclusions by explicit one-loop calculations for  muonic hydrogen. Unlike the ordinary hydrogen energy levels in muonic hydrogen even in the nonrecoil approximation depend on two scales, the electron and muon masses, and from a diagrammatic point of view it is not immediately clear why the trace diagrams generate correct results for the muonic hydrogen energy levels. We calculate the trace diagrams for muonic hydrogen in the one-loop approximation, show that they reproduce well known one-loop results for muonic hydrogen energy levels, and  explain why this happens.

\section{One-loop EMT trace diagrams}

In QED with two massive fermion flavors (electron and muon) the trace of EMT has the form

\beq \label{anomtrac}
{T_0^\mu}_\mu=[{T^\mu}_\mu]_R=(1+\gamma^e_m)[\bar\psi_e m_e\psi_e]_R+(1+\gamma^\mu_m)[\bar\psi_\mu m_\mu\psi_\mu]_R+\frac{\beta(e)}{2e}[F^2]_R,
\eeq

\noindent
where the one-loop $\beta$-function, $\beta(e)=\beta^e(e)+\beta^\mu(e)$,  is a sum of the electron and muon contributions  $\beta^e(e)/2e=\beta^\mu(e)/2e=\alpha/(6\pi)$, mass anomalous dimensions $\gamma^e_m=\gamma^\mu_m=3\alpha/(2\pi)$, $m_{e(\mu)}$ is the mass of the respective fermion, not the mass of a particle or bound state under consideration. The conserved EMT operator is not renormalized, so $T_0^{\mu\nu}$ in terms of bare fields (from which the bare (total) Lagrangian is constructed) coincides with the renormalized EMT $[T^{\mu\nu}]_R$\footnote{We use mass-shell renormalization scheme.}, which generates renormalized (UV finite) Green functions with renormalized fields $\psi$. Notice, that due to the scale anomaly the EMT trace is nonzero even in QED and QCD with massless leptons (quarks). 

We calculate the matrix element of the EMT trace in \eq{anomtrac} in the Furry picture \cite{Furry:1951bef,Sapirstein:1990gn,Weinberg:1995mt}, where the role of the free muon propagator plays the  Dirac-Coulomb Green's function

\beq
G=\frac{i}{p_0-\bm\alpha\cdot \bm p-\beta (m_\mu-i\epsilon)-V}\gamma_0
\equiv\frac{i}{E-H}\gamma_0,
\eeq

\noindent
$V=-Z\alpha/r$ is the Coulomb potential and $i$ in the numerator is included for consistency with the free Feynman propagator.

In terms of eigenfunctions the propagator has the form

\beq \label{dircoulgrfun}
G(E, \bm r,\bm r')
=\left\langle\bm r\left|\frac{i}{E-H}\gamma_0\right|\bm r'\right\rangle
=i\left[\sum_{n} \frac{\psi^{(+)}_n(\bm r) \bar  \psi^{(+)}_n(\bm r')}{E-E_n+i\epsilon}+\sum_{n} \frac{ \psi^{(-)}_n(\bm r) \bar\psi^ {(-)}_n(\bm r')}{E+E_n-i\epsilon}\right],
\eeq

\noindent
where summation goes over all states of discrete and continuous spectrum, $\psi^{(+)}_n(\bm r)$ and $\psi^{(-)}_n(\bm r)$ are eigenfunctions of the Dirac Hamiltonian in the external Coulomb field with positive and negative energies, respectively. These eigenfunctions are  normalized to one with the integration measure $\int d^3r$. 

The one-loop muonic hydrogen matrix element in \eq{traceqmass} in the Furry picture in terms of the renormalized fields has the form (see Appendix \ref{renormsal} for notation)

\beq
\label{anomtrac2}
\begin{split}
T\approx& \int d^3r\langle \mu|[m_e-\delta m_e+m\gamma_m^e(e)+m_e\delta Z_{2e}]\bar\psi_e(\bm r)\psi_e(\bm r)+\frac{\beta^e(e)}{2e}F^2(\bm r)\\
&+[m_\mu-\delta m_\mu+m\gamma_m^\mu(e)+m_\mu\delta Z_{2\mu}]\bar\psi_\mu(\bm r)\psi_\mu(\bm r)+\frac{\beta^\mu(e)}{2e}F^2(\bm r)|\mu\rangle,
\end{split}
\eeq

\noindent 
where $|\mu\rangle$ is the muon bound state in the external Coulomb field.

\begin{figure}[h!]
\begin{center}
\includegraphics[width=12cm]{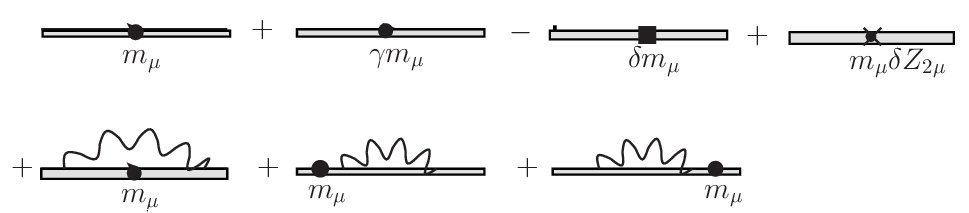}
\end{center}
\caption{Self-energy type trace Lamb shift diagrams.}
\label{hydemtdigse}
\end{figure}

\begin{figure}[h!]
\begin{center}
\includegraphics[height=2.5cm]{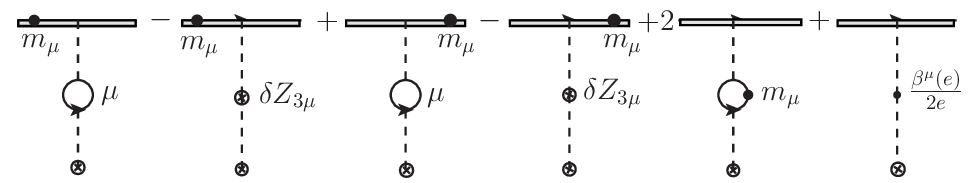}
\end{center}
\caption{Muon vacuum polarization type trace Lamb shift diagrams.}
\label{hydemtdigfp}
\end{figure}

\begin{figure}[h!]
\begin{center}
\includegraphics[height=2.5 cm]
{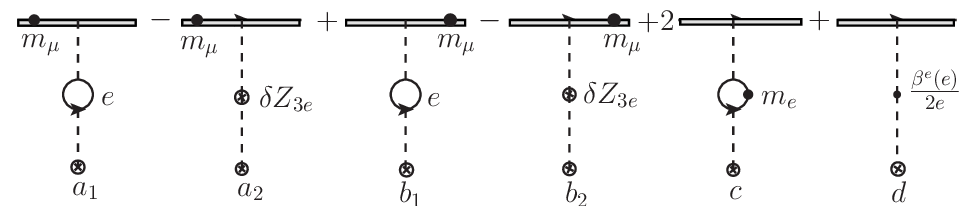}
\end{center}
\caption{Electron vacuum polarization type trace Lamb shift diagrams.}
\label{vacpolmisel}
\end{figure}

\begin{figure}[h!]
\begin{center}
\includegraphics[height=2.5cm]
{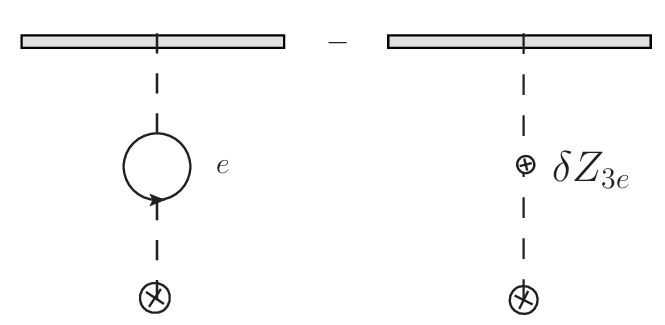}
\end{center}
\caption{Classical electron polarization Lamb shift  diagrams in muonic hydrogen.}
\label{lshdfpvpe}
\end{figure}

There are two sets of  the tree and one-loop diagrams for the matrix element in \eq{anomtrac2}: self-energy type diagrams in Fig.~\ref{hydemtdigse} and vacuum polarization type diagrams in Fig.~\ref{hydemtdigfp}  and in Fig.~\ref{vacpolmisel}. The trace diagrams in Fig.~\ref{hydemtdigse} and Fig.~\ref{hydemtdigfp} differ from the respective diagrams in electronic hydrogen only be the substitution $m_e\to m_\mu$. In that case the sum of diagrams in Fig.~\ref{hydemtdigse} and Fig.~\ref{hydemtdigfp} reproduces the Dirac energy and one-loop Lamb shift contributions to the energy levels \cite{Eides:2024tzh}. This is not the case for muonic hydrogen. The new element in muonic hydrogen is connected with the  electron polarization contributions in Fig.~\ref{vacpolmisel}. They do not coincide with the well known muonic hydrogen electron vacuum polarization diagrams in Fig.~\ref{lshdfpvpe}. According to \eq{traceqmass} and the trace theorem both sets of diagrams lead to the same results, and we would like to understand what properties of diagrams are responsible for this coincidence. We considered similar diagrams for electronic hydrogen in \cite{Eides:2024tzh} and demonstrated that the trace diagrams arise as logarithmic mass derivatives of the diagrams in Fig.~\ref{lshdfpvpe}. The only subtlety was that one needs to remember to differentiate the state vectors in the matrix elements. The energy levels in the electronic hydrogen are linear in the electron mass and therefore logarithmic mass derivative of an energy level (matrix element of the sum of the trace diagrams) coincides with the energy level itself. We also proved this by directly calculating the sum of the EMT trace diagrams. 

Due to presence of two different mass scales ($m_e$ and $m_\mu$) the logic based on linearity is not directly applicable for muonic hydrogen. Below we generalize the linearity argument for equality of the standard Lamb shift diagrams in muonic hydrogen and trace diagrams in Fig.~\ref{hydemtdigse}, Fig.~\ref{hydemtdigfp} and   Fig.~\ref{vacpolmisel}.  We also directly calculate contribution of the diagrams in Fig.~\ref{vacpolmisel} to the one-loop Lamb shift.

\section{Multiscale Problems\label{multiscc}}

The equality of an energy level and its logarithmic mass derivative does not hold beyond the nonrecoil approximation  even in electronic hydrogen and in the bound state problems with multiple mass scales, like muonic hydrogen. Then arises a natural question how the matrix element of the EMT trace in \eq{traceqmass} reproduces the bound state energy levels in the problem with multiple scales. We address this problem in the case of muonic hydrogen, which is one of the simplest bound states with two mass scales. Below we explicitly calculate the contribution of the diagrams in Fig.~\ref{vacpolmisel} and show that it coincides with the contribution of the diagrams in Fig.~\ref{lshdfpvpe}. Combined with the results for the diagrams in Fig.~\ref{hydemtdigse} and Fig.~\ref{hydemtdigfp} in \cite{Eides:2024tzh} (after the natural substitution $m_e\to m_\mu$ there), this proves  that the matrix element in \eq{traceqmass}  and \eq{anomtrac2} generates a correct one-loop contribution to the energy levels in muonic hydrogen.

Now we would like to address a more general case of energy levels of a bound state with a few independent  mass parameters $m_i$, $i=1,2,\ldots,k$. Energy of any state characterized by a multi-index $n$ has dimension of mass, so even in a multiscale problem $E_n(m_1,m_2,\ldots, m_k)$ is a homogeneous function of the first degree, i.e., satisfies the condition

\beq
E_n(\alpha m_1,\alpha m_2,\ldots, \alpha m_k)=\alpha  E_n(m_1,m_2,\ldots, m_k),
\eeq

\noindent
where $\alpha$ is an arbitrary parameter.

\noindent
Then

\beq
\frac{d}{d\alpha}E_n(\alpha m_1,\alpha m_2,\ldots, \alpha m_k)=E_n(m_1,m_2,\ldots, m_k).
\eeq

\noindent
On the other hand

\beq
\frac{d}{d\alpha}E_n(\alpha m_1,\alpha m_2,\ldots, \alpha m_k)
=\sum_{i=1}^{i=k}\frac{\partial E_n(\alpha m_1,\alpha m_2,\ldots, \alpha m_k)}{\partial(\alpha m_i)}m_i,
\eeq

\noindent
and at $\alpha=1$ we obtain the Euler's homogeneous function theorem

\beq \label{genhomrel}
E_n(m_1,m_2,\ldots, m_k)
=\sum_{i=1}^{i=k}\frac{\partial E_n(m_1, m_2,\ldots,  m_k)}{\partial m_i}m_i.
\eeq

\noindent
We see that the sum of the logarithmic mass partial derivatives of an energy level in a multiscale problem coincides with the energy level itself. Diagrammatically this means that the sum of diagrams for an energy level coincides with the sum of another set of diagrams obtained from the first set by logarithmic differentiation with respect to all mass parameters. On the basis of our experience with the free electron \cite{Eides:2023uox} and ordinary hydrogen \cite{Eides:2024tzh} we expect that logarithmic mass differentiation of the standard Lamb shift diagrams generates diagrams for the trace.

\begin{figure}[h!]
\begin{center}
\includegraphics[height=3 cm]
{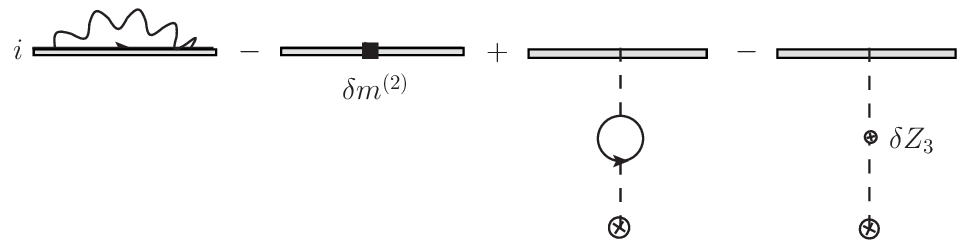}
\end{center}
\caption{Classical one-loop Lamb shift diagrams in electronic hydrogen.}
\label{lshdfp}
\end{figure}

Let us consider muonic hydrogen as a simple case of a multiscale problem. The classical one-loop Lamb shift diagrams in muonic hydrogen are in Fig.~\ref{lshdfpvpe} and Fig.~\ref{lshdfp}, where in the last figure all fermion lines are muonic. In electronic hydrogen only the diagrams in  Fig.~\ref{lshdfp} contribute to the one-loop Lamb shift. We have demonstrated  in \cite{Eides:2024tzh} that after logarithmic mass differentiation these diagrams generate diagrams in Fig.~\ref{hydemtdigse} and Fig.~\ref{hydemtdigfp} with all electronic fermion lines. We have also calculated the sum of the diagrams in Fig.~\ref{hydemtdigse} and Fig.~\ref{hydemtdigfp} for the electronic hydrogen and checked that it reproduces the one-loop Lamb shift. After rescaling $m_e\to m_\mu$ these results hold for muonic hydrogen, in other words the one-loop contribution of the diagrams in Fig.~\ref{lshdfp} with all muon lines coincides with the sum of the diagrams in Fig.~\ref{hydemtdigse} and Fig.~\ref{hydemtdigfp}.

The leading classical one-loop contribution  to the Lamb shift in muonic hydrogen is due to the diagrams in Fig.~\ref{lshdfpvpe}. By the same mechanism as considered in \cite{Eides:2024tzh} the logarithmic muon mass derivative of the state vectors not shown explicitly in the diagrams generates the first four diagrams in Fig.~\ref{vacpolmisel} with the muon mass insertions in the muon line. The logarithmic electron mass derivative differentiates the electron loop and the electron contribution to the renormalization constant and generates two last diagrams in Fig.~\ref{vacpolmisel}. Due to the general relationship in \eq{genhomrel} the contributions of the diagrams in Fig.~\ref{lshdfpvpe} and Fig.~\ref{vacpolmisel} should coincide. We check this general prediction by  direct calculations in the next section.

\section{Diagrams with electron polarization}

\subsection{One-loop electron polarization loop contribution to the Lamb shift}

The classical electron polarization loop contribution to the Lamb shift in muonic hydrogen  is given by the diagrams  in Fig.~\ref{lshdfpvpe}.  The leading one-loop contribution of these diagrams is of order $\alpha(Z\alpha)^2m_r$ (see below), and, hence, it is sufficient to use the free electron propagator in the polarization loop, account for the binding effects in this loop generates contributions of higher orders in $Z\alpha$. Unlike ordinary hydrogen in muonic hydrogen contribution of $P$-level is not parametrically suppressed in comparison with the contributions of $S$ levels, so we need to calculate both to obtain the Lamb shift

The field Hamiltonian $H^e_{int}=\int d^3 x{\cal H}^e_{int}=e\int d^3 x\bar\psi_\mu\gamma_0\psi_\mu A^0_{ext}(x)$  describes interaction of the static external Coulomb field  with the muon. One-loop corrected static Coulomb field in Fig.~\ref{lshdfpvpe} has the form (see, e.g., \cite{Berestetskii:1982qgu})

\beq \label{extpotco}
A^0_{ext,e~loop}(\bm r)=-\frac{Ze}{4\pi r}\frac{2\alpha}{3\pi}\int_1^\infty d\zeta e^{-2\rho\beta\zeta}\left(1+\frac{1}{2\zeta^2}\right)\frac{\sqrt{\zeta^2-1}}{\zeta^2},
\eeq

\noindent 
where $\rho=m_r Z\alpha r$, $m_r=m_\mu m_p/(m_\mu+m_p)$ is the reduced muon-proton mass and $\beta=m_e/(m_r Z\alpha)$. 

The respective leading nonrelativistic contribution to the Lamb shift is

\beq \label{matrlelhintexf}
\begin{split}
\Delta& E^e_{VP}(n,\ell)=\langle n\ell|{H}^e_{int}|n\ell\rangle
=e\int d^3r\langle n\ell|\bar\psi_\mu(\bm r) \gamma_0\psi_\mu(\bm r)  A^0_{ext,e~loop}(\bm r)|n\ell\rangle\\
&= -\frac{2(Z\alpha)\alpha}{3\pi}\int \frac{d^3r}{r}\psi^\dagger_{n\ell}(\bm r)
\psi_{n\ell}(\bm r)\int_1^\infty d\zeta e^{-2\rho\beta\zeta}\left(1+\frac{1}{2\zeta^2}\right)\frac{\sqrt{\zeta^2-1}}{\zeta^2}\\
&
= -\frac{8\alpha(Z\alpha)^2m_r}{3\pi n^3}
\int_0^\infty \rho d\rho f^2_{n\ell}(\rho_n)
\int_1^\infty d\zeta e^{-2\rho\beta\zeta}\left(1+\frac{1}{2\zeta^2}\right)\frac{\sqrt{\zeta^2-1}}{\zeta^2},
\end{split}
\eeq

\noindent 
where $f_{n\ell}(\rho_n)$ is defined in Appendix \ref{schrcoul}. 

Numerically for $n=2$ we obtain

\beq \label{stnasvacpolmu}
\begin{split}
&\Delta E^e_{VP}(n=2,\ell=0)=-219.5839\ldots~\mathrm{meV},\quad \Delta E^e_{VP}(n=2,\ell=1)=-14.5765\ldots~\mathrm{meV},\\
& 
\Delta E^e_{VP}(n=2,\ell=0)-\Delta E^e_{VP}(n=2,\ell=1)=-205.0073\ldots~\mathrm{meV}.
\end{split}
\eeq

\subsection{Calculations of the trace diagrams in Fig.~\ref{vacpolmisel}}

In the case of electronic hydrogen calculations of the diagrams in Fig.~\ref{vacpolmisel} were simplified by the fact that the characteristic scale of exchanged momenta $m_eZ\alpha$  is much smaller than momenta of order $m_e$ in the polarization loop. Due to this observation it was sufficient to account only for the leading term in the small momenta expansion of the polarization loop what allowed us to conclude before calculations that the sum of the last two diagrams in Fig.~\ref{vacpolmisel} is equal  $-2E^e_{VP}$ \cite{Eides:2024tzh}, where $E^e_{VP}$ is the contribution of the diagrams in Fig.~\ref{lshdfpvpe}. Moreover, we also concluded before calculation that the contributions of the first four sidewise diagrams in  Fig.~\ref{vacpolmisel} is equal $3E^e_{VP}$, and therefore sum of all diagrams in Fig.~\ref{hydemtdigfp} reproduces the standard contribution of the  diagrams in Fig.~\ref{lshdfpvpe} in electronic hydrogen. All these conclusions were confirmed in \cite{Eides:2024tzh} by direct calculations.

The exchanged momenta in muonic hydrogen are of order $m_\mu Z\alpha$ and the electron polarization loop momenta of order $m_e$ have comparable magnitude and we cannot use low momenta expansion of the electronic loop. Hence, the diagrams with the electronic loops in Fig.~\ref{vacpolmisel} in muonic hydrogen should be calculated from scratch. Like in ordinary hydrogen we expect that diagram $(c)$ in Fig.~\ref{vacpolmisel} with the scalar vertex insertion in the electron loop, which is equal the logarithmic derivative of the unrenormalized diagram with the electron loop, will generate a term which will cancel contribution of diagram $(d)$ with the $\beta_e$-function in Fig.~\ref{vacpolmisel},  which arises from the logarithmic derivative of the contribution to $\delta Z_{3e}$ of the electron loop.

\subsubsection{Matrix element of $m_\mu\bar\psi_\mu\psi_\mu$  with sidewise insertion of the polarization loop}

The first four external field diagrams in Fig.~\ref{vacpolmisel} arise as the one-loop perturbation theory corrections to the matrix element of the scalar vertex $m_\mu\bar\psi_\mu\psi_\mu$ in \eq{anomtrac}. The electron loops in these diagrams are renormalized by the two diagrams with $\delta Z_{3e}$, which are due to the Lagrangian counterterm.  We use the standard one-loop expression for the renormalized electron polarization loop for their calculation. The contributions to the energy shift of the diagrams $(a_1)$ and $(b_1)$ in Fig.~\ref{vacpolmisel} with the renormalized electron one-loop insertion in the Coulomb photon  are equal and can be written in the form

\beq \label{sidewisepolm}
\Delta E_a=\Delta E_b
=e\int d^3rd^3r'\psi_{n}(\bm r)A^0_{ext,e~loop}(\bm r)[-iG_r(\bm r,\bm r',E_n)]m\gamma_0\psi_{n}(\bm r'),
\eeq

\noindent
where $A^0_{ext,e~loop}(\bm r)$ is defined in \eq{extpotco}, $\psi_{n}(\bm r)$ is the Dirac-Coulomb bound state wave function and $G_r(E, \bm r,\bm r')$ is the reduced Dirac-Coulomb Green's function (compare \eq{dircoulgrfun})

\beq \label{redgrefunc}
G_r(E, \bm r,\bm r')
=\left\langle\bm r\left|\left(\frac{i}{E-H}\right)'\gamma_0\right|\bm r'\right\rangle
=\left\langle\bm r\left|\sum_{k\neq n}\frac{i|k\rangle\langle k|}{E-E_k}\gamma_0\right|\bm r'\right\rangle.
\eeq

We define a perturbed wave function $\widetilde{\psi}_{n}$  as 

\beq \label{sidewisepolm3}
\widetilde{\psi}_{n}(\bm r) = \int d^3r' 
 [-iG_r(\bm r,\bm r',E_n)]m\gamma_0\psi_{n}(\bm r')
\eeq

\noindent 
Then \eq{sidewisepolm} acquires the form 

\beq \label{sidewisepolm2}
\Delta E_a=\Delta E_b
=e\int d^3r\psi_{n}(\bm r)A^0_{ext,e~loop}(\bm r) \widetilde{\psi}_{n}(\bm r).
\eeq

The perturbed wave function  $\widetilde{\psi}_{n}(\bm r)$ can be calculated analytically 
in a closed form 
with the help of the virial relationships derived in \cite{shabaev1991,shabaev2002}. 
The explicit relativistic formulas for the four-component Dirac wave functions
were reported 
\footnote{
We use the opportunity to correct two misprints in Ref.~\cite{Eides:2024tzh}. First, in the second line in Eq.~(C5) the third term should be $G$, not $F$. Second, the term $Z\alpha$ in the second line of Eq.~(C21) should have an opposite sign, as immediately follows from Eq.~(C20). 
}
in Appendix C of Ref.~\cite{Eides:2024tzh}. 
In the present work we perform calculations only in the lowest order in the parameter $Z\alpha$. 
It is thus sufficient to use the nonrelativistic limit of the perturbed wave function,
which has the simple form

\beq \label{sidewisepolm4}
\widetilde{\psi}_{n\ell }(\bm r) = \left( \frac32 + r \frac{d}{dr}\right)\psi_{n\ell }(\bm r),
\eeq

\noindent
and respectively 

\beq
\widetilde f_{n\ell}(\rho_n)=\left( \frac32 + r \frac{d}{dr}\right)f_{n\ell}(\rho_n).
\eeq

\noindent
Then the contribution to the energy shift  in \eq{sidewisepolm2} can be written in the form   similar to \eq{matrlelhintexf}

\beq \label{pertformwf}
\begin{split}
\Delta E_{a,b}(n,\ell)&=
-\frac{2(Z\alpha)\alpha}{3\pi}\int \frac{d^3r}{r}\psi^\dagger_{n\ell}(\bm r)
\widetilde \psi_{n\ell}(\bm r)\int_1^\infty d\zeta e^{-2\rho\beta\zeta}\left(1+\frac{1}{2\zeta^2}\right)\frac{\sqrt{\zeta^2-1}}{\zeta^2}\\
&
= -\frac{8\alpha(Z\alpha)^2m_r}{3\pi n^3}
\int_0^\infty \rho d\rho f_{n\ell}(\rho_n)\widetilde f_{n\ell}(\rho_n)
\int_1^\infty d\zeta e^{-2\rho\beta\zeta}\left(1+\frac{1}{2\zeta^2}\right)\frac{\sqrt{\zeta^2-1}}{\zeta^2}.
\end{split}
\eeq

Numerically for $n=2$ we obtain 

\beq \label{musideab}
\begin{split}
&\Delta E_{a,b}(n=2,\ell=0)=-235.5288\ldots~\mathrm{meV},\quad \Delta E_{a,b}(n=2,\ell=1)=-27.2887\ldots~\mathrm{meV},\\
&
\Delta E_{a,b}(n=2,\ell=0)-\Delta E_{a,b}(n=2,\ell=1)=-208.2401\ldots~\mathrm{meV}.
\end{split}
\eeq

\subsubsection{Matrix element of the scalar vertex $m_e\bar \psi_e\psi_e$ insertion in the polarization loop}

Two identical diagrams $(c)$  in Fig.~\ref{vacpolmisel} with insertion of the scalar vertex $m_e\bar\psi_e\psi_e$ in the electron polarization loop arise as one-loop radiative corrections to the electron mass term in the EMT trace, see \eq{anomtrac}. Contribution to the energy shift from the sum of these identical diagrams has the form

\beq \label{edlec}
\Delta E_c(n,\ell)=2e\int d^3r\langle n\ell|m_e\bar\psi_e(\bm r)\psi_e(\bm r)|n\ell\rangle
=2e\int d^3r\psi^\dagger_{n\ell}(\bm r)\psi_{n\ell}(\bm r)A^0_m(\bm r),
\eeq

\noindent
where $A^0_m(\bm r)$ is the radiative correction to the Coulomb potential due to one of the two diagrams $(c)$ in Fig.~\ref{vacpolmisel}. 

\begin{figure}[h!]
\begin{center}
\includegraphics[height=3.cm]
{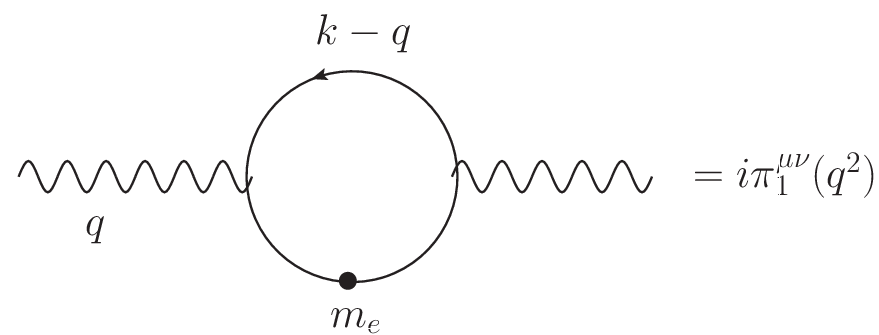}
\end{center}
\caption{Polarization loop with the scalar vertex insertion $m_e$.}
\label{polminsl}
\end{figure}

Field $A^0_m(\bm r)$ is determined by the electron polarization loop $i\pi_1^{\mu\nu}(q)$ with the  mass insertion $m_e$ in Fig.~\ref{polminsl}, which is defined by the Feynman integral

\beq \label{ipi1polm}
i\pi_1^{\mu\nu}(q)=(-ie)^2(-1)m_e\int \frac{d^4k}{(2\pi)^4}Tr\left[\gamma^\mu
\left(\frac{i}{\slashed k-m_e+i\epsilon}\right)^2\gamma^\nu
\frac{i}{\slashed k-\slashed q-m_e+i\epsilon}\right].
\eeq

\noindent
Naively this integral is linearly divergent, but due to gauge invariance $i\pi_1^{\mu\nu}(q)=i(g^{\mu\nu}q^2-q^\mu q^\nu)\pi_1(q^2)$ and the remaining integral is convergent and does not require  any new counterterm. This is unlike the case of the standard polarization loop, where even after account for gauge invariance the logarithmic divergence survives and requires a counterterm. Calculating $\pi_1(q^2)$ we obtain (see eq.(26) in \cite{Eides:2024tzh})

\beq \label{lowmomexppm}
\begin{split}
\pi_1(q^2)&=\frac{e^2m_e^2i}{\pi^2}\frac{1-\frac{4m_e^2 \tanh ^{-1}\left(\sqrt{\frac{-q^2}{4m_e^2-q^2}}\right)}{\sqrt{-q^2 (4 m_e^2-q^2)}}}{-2q^2}_{|q^2=-\bm q^2, \bm q^2/m_e^2\to0}\\
&\to \frac{2\alpha i}{\pi}\left(\frac{1}{6}-\frac{\bm q^2}{30m_e^2}+\frac{\bm q^4}{140m_e^4}+\ldots\right).
\end{split}
\eeq

\noindent
Next we separate the leading term in the small momentum expansion

\beq \label{pi1qq}
\pi_1(q^2)
=\frac{\alpha}{3\pi}i +i\frac{e^2}{\pi^2}\left[m_e^2\frac{1-\frac{4m_e^2 \tanh ^{-1}\left(\sqrt{\frac{\bm q^2}{4m_e^2+\bm q^2}}\right)}{\sqrt{\bm q^2 (4 m_e^2+\bm q^2)}}}{2\bm q^2}
-\frac{1}{12}
\right]
=\frac{\alpha}{3\pi}i +i\frac{e^2}{\pi^2}\widetilde\pi_1(-\bm q^2).
\eeq

\noindent
We separated the first term because, as we will see below, it will exactly cancel with the contribution of the anomalous EMT trace term $\beta/(2e)F^2$ in the diagram $(d)$ in Fig.~\ref{vacpolmisel}.

In terms of $\widetilde\pi_1(-\bm q^2)$ the radiatively corrected Coulomb field $A^0_m(\bm r)$ has the form

\beq \label{extmodpi1}
A^0_m(\bm r)=-Ze\int\frac{d^3q}{(2\pi)^3}e^{i\bm q\cdot\bm r}\frac{1}{\bm q^2}\left(\frac{\alpha}{3\pi} +\frac{e^2}{\pi^2}\widetilde\pi_1(-\bm q^2)\right)
=-\frac{Ze\alpha}{12\pi^2 r} -\frac{Ze^3}{\pi^2}\int\frac{d^3q}{(2\pi)^3}e^{i\bm q\cdot\bm r}\frac{\widetilde\pi_1(-\bm q^2)}{\bm q^2}.
\eeq

\noindent
Next we plug this expression in \eq{edlec}, and arrive at the contribution to the energy level

\beq \label{incompldic}
\begin{split}
&\Delta E_c(n,\ell)=
2e\int d^3r\psi^\dagger_{n\ell}\psi_{n\ell}(\bm r)A^0_m(\bm r)\\
&
=-\frac{2\alpha (Z\alpha)}{3\pi}\int d^3r\psi^\dagger_{n\ell}\frac{1}{r}\psi_{n\ell}(\bm r)
-32\alpha(Z\alpha)\int d^3r\psi^\dagger_{n\ell}\psi_{n\ell}(\bm r)\int\frac{d^3q}{(2\pi)^3}e^{i\bm q\cdot\bm r}\frac{\widetilde\pi_1(-\bm q^2)}{\bm q^2}\\
&
\equiv \Delta E_{c1}+\Delta E_{c2}.
\end{split}
\eeq

We will  discuss $\Delta E_{c1}$ below together with the contribution of the  diagram $(d)$ in Fig.~\ref{vacpolmisel}. To calculate $\Delta E_{c2}$ we plug  the explicit expression for $\widetilde\pi_1(-\bm q^2)$ from \eq{pi1qq} in the  integral in \eq{incompldic} and rescale the integration variable $q=m_rZ\alpha \widetilde q$. Then we obtain $\Delta E_{c2}(n\ell)$ in the form

\beq  \label{nainloopem}
\Delta E_{c2}(n\ell)
=-\frac{64\alpha(Z\alpha)^2m_r}{\pi^2n^3}\int d\rho \rho f^2_{n\ell}(\rho_n)
\int_0^\infty d\widetilde q\frac{\sin(\widetilde { q}\rho)}{\widetilde { q}}
\left[\beta^2\frac{1-\frac{4\beta^2 \tanh ^{-1}\left(\sqrt{\frac{\widetilde{\bm q}^2}{4\beta^2+\widetilde{\bm q}^2}}\right)}{\sqrt{\widetilde{\bm q}^2 (4 \beta^2+\widetilde{\bm q}^2)}}}{2\widetilde{\bm q}^2}
-\frac{1}{12}
\right] 
\eeq

Numerically for $n=2$ we obtain

\beq
\begin{split} \label{muelpolminsv}
&\Delta E_{c2}(n=2,\ell=0)=251.4737\ldots~\mathrm{meV},\quad \Delta E_{c2}(n=2,\ell=1)=40.0008\ldots~\mathrm{meV},\\
&
\Delta E_{c2}(n=2,\ell=0)-\Delta E^e_{c2}(n=2,\ell=1)=211.4729\ldots~\mathrm{meV}.
\end{split}
\eeq

Notice that as in electronic hydrogen this contribution has an opposite sign to the vacuum polarization result in \eq{stnasvacpolmu} (in that case it was $-2\Delta E^e_{VP}$, but now we do not expect to get the same factor 2).

\subsubsection{Matrix element of the anomalous term $(\beta_e/2e)F^2$ insertion in the Coulomb photon}

Diagram $(d)$ in Fig.~\ref{vacpolmisel} arises as matrix element of the electron loop contribution to the anomalous EMT term  $(\beta_e(e)/2e)F^2$ in \eq{anomtrac}. This diagram  is similar to diagram $(c)$ in Fig.~\ref{vacpolmisel}, the only difference is that instead of insertion of the term $i\pi_1^{\mu\nu}$ in the external Coulomb propagator as in \eq{extmodpi1}, we now insert the two-prong vertex $(\beta_e/2e)F^2$. In momentum space it has the form $4(\beta_e/(2e))(g_{\mu\nu}q^2-q_\mu q_\nu)$, which reduces to insertion of $4(\beta_e/(2e))$ in the Coulomb line. We use $4\beta_e(e)/2e=2\alpha/3\pi$ and obtain the leading nonrelativistic contribution of the diagram $(d)$ in Fig.~\ref{vacpolmisel} 

\beq 
\Delta E_{d}=\frac{2\alpha(Z\alpha)}{3}\int d^3 r\psi^\dagger_{nl}(\bm r)\psi_{nl}(\bm r).
\eeq

\noindent
This contribution has the same magnitude, but an opposite sign as the contribution  $\Delta E_{c1}$ in \eq{incompldic}

\beq \label{anommamffelneq}
\Delta E_{d}=-\Delta E_{c1}.
\eeq

\noindent
Cancellation between the contribution of the anomalous EMT trace term in Fig.~\ref{vacpolmisel} $(d)$ and the term  $\Delta E_{c1}$ from diagram $(c)$ in Fig.~\ref{vacpolmisel} happens not by an accident. Recall that the diagram $(c)$ arises as a logarithmic electron mass derivative of the unrenormalized electron loop in Fig.~\ref{lshdfpvpe}. This diagram implicitly contains a logarithmically divergent term $\delta Z_{3e}$ which becomes a finite constant contribution to the electron loop in Fig.~\ref{vacpolmisel} after differentiation and the leading term in the low-momentum expansion in \eq{lowmomexppm}. But 

\beq \label{derpolrens}
m\frac{d\delta Z_3}{dm}=-\mu\frac{d\delta Z_3}{d\mu}
=\frac{2\beta(e)}{e}\approx \frac{2\alpha}{3\pi}.
\eeq

\noindent
The first equality on the RHS holds because the counterterm $\delta Z_3=\Pi_{reg}^{(2)}(0)$ is linear in $\ln(\mu/m)$, see \eq{conutertermpold}. We see that cancellation of the anomalous contribution of the diagram $(d)$ in  Fig.~\ref{vacpolmisel} is due to the definition of the $\beta$-function.

Let us mention that the cancellation is due exclusively to the definition of the $\beta$-function and holds also beyond the nonrelativistic approximation for the wave functions.


\section{Summary}

We derived explicit formulae (see \eq{matrlelhintexf}, \eq{pertformwf} and \eq{nainloopem}) for the one-loop nonrelativistic contributions to the Lamb shift in muonic hydrogen from the polarization type EMT trace diagrams with the electron loop.  We used these formulae to calculate numerically Lamb shift contributions of the trace  diagrams for $n=2$. To obtain the total contribution of the electron polarization type diagrams in Fig.~\ref{vacpolmisel} for $n=2$ we sum contribution is \eq{musideab} and \eq{muelpolminsv}

\beq
\begin{split}
&\Delta E(n=2,\ell=0)=2\Delta E_{a,b}(n=2,\ell=0)+\Delta E_{c2}(n=2,\ell=0)=-219.5839\ldots~\mathrm{meV}, \\
&
\Delta E(n=2,\ell=1)=2\Delta E_{a,b}(n=2,\ell 1)+\Delta E^e_{c2}(n=2,\ell=1)=-14.5765\ldots~\mathrm{meV},\\
&
\Delta E(n=2,\ell=0)-\Delta E(n=2,\ell=1)=-205.0073\dots~\mathrm{meV}.
\end{split}
\eeq

\noindent 
These results are in full agreement with the standard electron loop polarization contribution  in \eq{stnasvacpolmu} from the diagrams in  Fig.~\ref{lshdfpvpe}.  Hence,  the sum of all one-loop EMT trace diagrams Fig.~\ref{hydemtdigse}, Fig.~\ref{hydemtdigfp} and Fig.~\ref{vacpolmisel} reproduces one-loop contribution to the Lamb shift in muonic hydrogen, as it should in accordance with the gauge and renormalization scheme independent \eq{traceqmass} and \eq{anomtrac2}.  

Technically, at the one-loop  level, this happens due two observations. First,  the energy of an energy level of a bound state is a homogeneous function of all independent mass parameters of the first degree. Then, due to the Euler's theorem, the sum of logarithmic mass derivatives with respect to all independent mass parameters coincides with the energy level itself. Second, logarithmic mass derivatives of the standard diagrams for an energy level generate the EMT trace diagrams. We demonstrated this last property considering one-loop corrections to energy levels in muonic hydrogen, but we expect this relationship between the diagrams to hold beyond the leading order as well. Taken together these two observation would provide an independent diagrammatic proof of \eq{traceqmass} for bound state with a few independent mass parameters.

\acknowledgments

M. Eides is grateful to Vladimir Pascalutsa for raising the problem of multiscale bound
states in the context of the EMT trace. Work of M. Eides was supported by the NSF grant PHY-2510100. 

\appendix

\section{One-loop renormalization constants\label{renormsal}}

We use dimensional regularization ($d=4-2\epsilon$) and mass-shell renormalization scheme. QED  Lagrangian in this scheme is

\beq
{\cal L}_0={\cal L}+\delta{\cal L}=-\frac{1}{4}F_0^2+\bar\psi_0(i\slashed\partial-m_0)\psi_0
-e_0\bar\psi_0\slashed A_0\psi_0,
\eeq

\noindent
where

\beq
\begin{split}
{\cal L}&=-\frac{1}{4}F^2+\bar\psi(i\slashed\partial-m)\psi
-\mu^\epsilon e\bar\psi\slashed A\psi, \\
%
%
\delta {\cal L}&=-\frac{1}{4}\delta Z_3F^2+\bar\psi(i\delta
Z_2\slashed\partial-\delta_m)\psi
-\mu^\epsilon e\delta Z_1\bar\psi\slashed A\psi.
\end{split}
\eeq

\noindent
The renormalization constants are defined as

\beq
\begin{split}
Z_1&=1+\delta Z_1,\quad Z_2=1+\delta Z_2, \quad Z_3=1+\delta Z_3, \quad
e_0=\mu^\epsilon Z_3^{-\frac{1}{2}}e,
\\
m_0&=mZ_mZ^{-1}_2,\quad
mZ_m=m(1+\delta Z_m)=m+\delta_m,\quad \delta m= m-m_0=m-mZ_mZ_2^{-1}.
\end{split}
\eeq

In the one-loop approximation\footnote{We use the Feynman gauge for calculations.}

\beq \label{polaregguidim}
\begin{split}
&\Pi_{reg}(q^2)=-\frac{2\alpha}{\pi}\int_0^1dxx(1-x)\left[\frac{1}{\tilde\epsilon}
+\ln\frac{\mu^2}{-x(1-x)q^2+m^2}
\right],\\
&\Sigma_{reg}(p)=\frac{\alpha}{2\pi}
\int_0^1dx\Biggl\{(2m-x\slashed p)
\left[\frac{1}{\tilde\epsilon}+
\ln\frac{\mu^2}{-x(1-x)p^2+x\lambda^2+(1-x)m^2}\right]
-(m-x\slashed p)\Biggr\},
\end{split}
\eeq

\noindent
where $\mu$ is the auxiliary dimensional regularization mass, $\lambda$ is the IR
photon mass, and $1/\tilde\epsilon={1}/{\epsilon}-\gamma+\ln(4\pi)$.

The one-loop counterterms  in the mass shell renormalization scheme are

\beq \label{conutertermpold}
\begin{split}
&\delta {Z_3}=\Pi_{reg}(0)=-\frac{\alpha}{3\pi}\left[\frac{1}{\tilde\epsilon}
+\ln\frac{\mu^2}{m^2}\right],\\
&m\delta Z_2-\delta_m=\Sigma_{reg}(m)
=\frac{3\alpha}{4\pi}m
\left[\frac{1}{\tilde\epsilon}+\ln\frac{\mu^2}{m^2}
+\frac{4}{3}\right]
\equiv \delta m,\\
&\delta Z_2=\Sigma'_{reg}(\slashed p=m)
=-\frac{\alpha}{4\pi}\left[\frac{1}{\tilde\epsilon}
+\ln\frac{\mu^2}{m^2}+2\ln\frac{\lambda^2}{m^2}+{4}\right],\\
&\delta Z_1=-\Lambda_{reg}(0)
=-\frac{\alpha}{4\pi}\left[\frac{1}{\tilde\epsilon}
+\ln\frac{\mu^2}{m^2}+2\ln\frac{\lambda^2}{m^2}+4\right],\\
&\delta_m\equiv\delta Z_m m=m\delta Z_2-\Sigma_{reg}(m)
=\frac{\alpha}{4\pi}m\left[-\frac{4 }{ \tilde\epsilon}- 2 \ln\left(\frac{\lambda^2}{m^2}\right)-4 \ln \left(\frac{\mu ^2}{m^2}\right)-8\right].
\end{split}
\eeq

\section{Schr\"odinger-Coulomb bound state wave functions\label{schrcoul}}

To specify notation and for completeness we collect below the Schr\"odinger-Coulomb wave functions for a particle with mass $m_r$ in the Coulomb potential $(-Z\alpha/r)$, see e.g., \cite{Landau:1991wop}. The normalized bound state wave function has the form\footnote{Unlike \cite{Landau:1991wop}, we use Laguerre polynomials as defined in MATHEMATICA.}

\beq 
\psi_{n\ell m}(\bm r)=Y_{\ell m}(\theta,\phi)R_{n\ell}(r),
\eeq

\noindent 
where  ($\rho_n=m_r Z\alpha r/n$) 

\beq \label{flnrnl}
R_{n\ell}(r)= 2\left(\frac{m_rZ\alpha}{n}\right)^\frac{3}{2}f_{n\ell}(\rho_n),
\eeq

\noindent 
and 

\beq
f_{n\ell}(\rho_n)=\sqrt{\frac{(n-l-1)!}{n(n+l)!}}
e^{-\rho_n}\left(2\rho_n\right)^l
{L^{2l+1}_{n-l-1}}\left(2\rho_n\right).
\eeq

\end{document}